\documentclass[12pt]{iopart}
\usepackage{amssymb,graphicx}

\newcommand{\dd}{\rmd}
\newcommand{\td}[2]{\frac{\dd #1}{\dd #2}}
\newcommand{\pd}[2]{\frac{\partial #1}{\partial #2}}
\newcommand{\fd}[2]{\frac{\delta #1}{\delta #2}}
\newcommand{\mean}[1]{\langle #1\rangle}

\newcommand{\IInt}[3]{\int_{#2}^{#3}\dd #1\;}
\newcommand{\eqref}[1]{(\ref{#1})}

\renewcommand{\vec}[1]{\mathbf #1}
\renewcommand{\mat}[1]{\mathsf #1}

\newcommand{\arsinh}{\mathrm{arsinh}}

\newcommand{\al}{\alpha}

\newcommand{\eps}{\varepsilon}
\newcommand{\kap}{\kappa}
\newcommand{\lam}{\lambda}

\newcommand{\sig}{\sigma}

\newcommand{\ra}{\rightarrow}

\newcommand{\sm}{s_\mathrm{m}}
\newcommand{\smm}{\mean{\dot s_\mathrm{m}}}
\newcommand{\vel}{\mean{\dot x}}
\newcommand{\ac}{\mathcal S}
\newcommand{\la}{\mathcal L}
\newcommand{\hs}{h_\mathrm{s}}
\newcommand{\tp}{\tau_0}
\newcommand{\fc}{f_\mathrm{c}}

\begin{document}

\title[LDF for entropy production: Optimal trajectory and role of
fluctuations]{Large deviation function for the entropy production: Optimal
  trajectory and role of fluctuations}

\author{Thomas Speck}
\address{Institut f\"ur Theoretische Physik II,
  Heinrich-Heine-Universit\"at D\"usseldorf, Universit\"atsstra\ss e 1, 40225
  D\"usseldorf, Germany}
\author{Andreas Engel}
\address{Universit\"at Oldenburg, Institut f\"ur Physik, 26111 Oldenburg,
  Germany}
\author{Udo Seifert}
\address{{II.} Institut f\"ur Theoretische Physik, Universit\"at
  Stuttgart, Pfaffenwaldring 57, 70550 Stuttgart, Germany}

\begin{abstract}
  We study the large deviation function for the entropy production rate in two
  driven one-dimensional systems: the asymmetric random walk on a discrete
  lattice and Brownian motion in a continuous periodic potential. We compare
  two approaches: the Donsker-Varadhan theory and the Freidlin-Wentzell
  theory. We show that the wings of the large deviation function are dominated
  by a single optimal trajectory: either in forward (positive rate) or in
  backward direction (negative rate). The joining of both branches at zero
  entropy production implies a non-differentiability and thus the appearance
  of a ``kink''. However, around zero entropy production many trajectories
  contribute and thus the kink is smeared out.
\end{abstract}

\maketitle
\tableofcontents


\section{Introduction}

Driving a system away from thermal equilibrium implies a non-vanishing entropy
production rate in the surrounding environment due to heat dissipation. For
mesoscopic systems, the energy thus exchanged with the environment is a
stochastic quantity, and so is the entropy $\sm$ produced along a single
trajectory. This observation forms the basis of stochastic
thermodynamics~\cite{seif12}. For a system driven into a non-equilibrium
steady state, the long-time limit of the corresponding probability
distribution $p(\sm,t)$ defines a large deviation function (LDF)
\begin{equation}
  \label{eq:ldf}
  h(\sig) \equiv \lim_{t\ra\infty} -\frac{\ln p(\sm,t)}{\smm t}
\end{equation}
with normalized entropy production rate
\begin{equation}
  \label{eq:sig}
  \sig \equiv \frac{\sm}{\smm t}.  
\end{equation}
Under very general assumptions, this large deviation function obeys the
Gallavotti-Cohen symmetry~\cite{gall95,kurc98,lebo99}
\begin{equation}
  \label{eq:gc}
  h(-\sig) = h(\sig) + \sig.
\end{equation}
Experimental tests of this relation have been performed in a variety of
systems, e.g., a single trapped colloidal particle~\cite{spec07,andr07} and a
mechanical oscillator~\cite{doua06}. Other large deviation functions also play
an important role for the study of non-equilibrium
systems~\cite{pani97,saha11,nemo11,touc13}.

\begin{figure}[t]
  \centering
  \includegraphics[width=.8\linewidth]{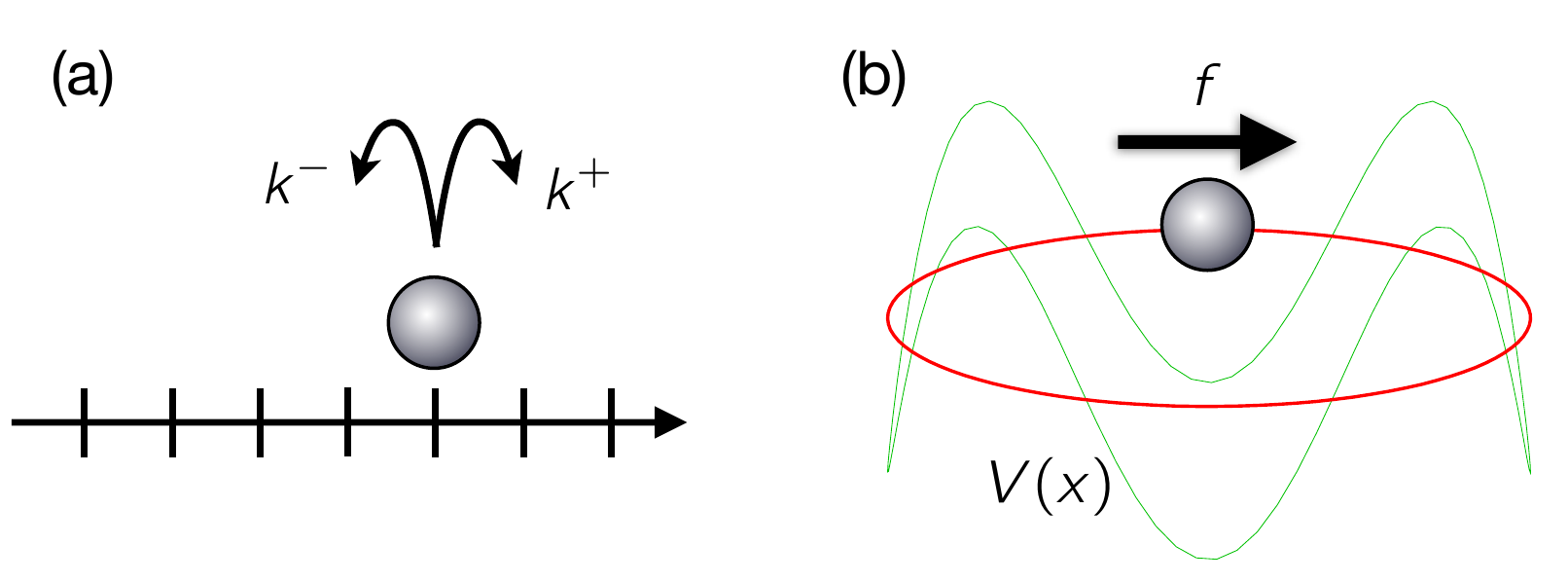}
  \caption{The two systems studied: (a)~the asymmetric random walk of a
    particle on a lattice with discrete positions. Different forward rate
    $k^+$ and backward rate $k^-$ can be interpreted as an effective driving
    force $f=\ln(k^+/k^-)$. (b)~The continuous analog is a particle moving in
    a periodic ring potential $V(x)$ and driven by a force $f$.}
  \label{fig:sys}
\end{figure}

We study two paradigmatic models for driven systems sketched in
Fig.~\ref{fig:sys}: the asymmetric random walk on a discrete lattice and
Brownian motion of a colloidal particle in a one-dimensional periodic
potential. The latter system has been studied extensively in the context of
the fluctuation-dissipation theorem~\cite{spec06,blic07,seif10,mehl10,gome11}.
In order to obtain explicit expressions for Eq.~\eqref{eq:ldf}, two main
approaches are discussed in the literature (for a comprehensive review and
references, see Ref.~\cite{touc09}). The first is the Donsker-Varadhan theory,
stating that the scaled cumulant generating function
\begin{equation}
  \label{eq:al}
  \al(\lam) \equiv \lim_{t\ra\infty}\frac{1}{t}\ln\mean{e^{\lam\sm}}
\end{equation}
is related to the LDF through the Fenchel-Legendre transformation
\begin{equation}
  \label{eq:legendre}
  h(\sig) = \sup_{\lam} \{ \lam\sig - \al(\lam)/\smm \}.
\end{equation}
This method has been used previously to obtain the LDF for the two systems we
study here: the asymmetric random walk~\cite{lebo99} and the continuous
periodic potential~\cite{mehl08}. The second approach due to Freidlin and
Wentzell~\cite{freidlin} focusses on paths instead of the scaled cumulant
generating function $\al(\lam)$. The probability of a path $x(\tau)$ obeys a
large deviation principle
\begin{equation}
  \label{eq:fw}
  P[x(\tau)] \sim e^{-J[x(\tau)]/\eps}
\end{equation}
in the limit of small noise $\eps\ra0$. The large deviation function $h(\sig)$
can then be obtained from the optimal trajectory $x_\ast(\tau)$ minimizing
$J[x(\tau)]$ using the contraction principle. While the Donsker-Varadhan
theory makes no assumptions on the noise strength and is thus more general, it
is also less intuitive. The Freidlin-Wentzell theory, on the other hand, is
more appealing to physical intuition since it allows a more direct insight.

The purpose of this work is two-fold: First, we provide a detailed study
comparing for two specific models the two approaches to the calculation of the
large deviation function. Based on the Freidlin-Wentzell approach, the second
purpose is to study the origin of a particular feature of the LDF: The
emergence of a ``kink'' around zero entropy production, an abrupt albeit
differentiable change of the slope~\cite{mehl08}. This ``kink'' has been
observed in distributions of the velocity (which is closely related to the
entropy production) in models for molecular motors~\cite{laco08}, and
experimentally for a tagged granular particle~\cite{kuma11}. Of course, this
``kink'' does not indicate a phase transition. Dynamic phase transitions are
indicated by a jump discontinuity of either the first or second derivative of
the scaled cumulant generating function $\al(\lam)$. Such transitions have
been found in interacting systems, e.g., for the totally asymmetric exclusion
process~\cite{gori11}, the symmetric exclusion process~\cite{leco12}, and for
models with kinetic constraints~\cite{spec11}.

Basically two theoretical explanations for the physical origin of the kink
have been proposed so far. Pleimling and coworkers~\cite{doro11,avra11} have
claimed that the kink is a generic feature of LDFs for the entropy production
that follows quite generally from the Gallavotti-Cohen symmetry
Eq.~\eqref{eq:gc}. But the relation Eq.~\eqref{eq:gc} only restricts the
anti-symmetric part, i.e., it allows us to write $h(\sig)=\hs(\sig)-\sig/2$
with $\hs(-\sig)=\hs(\sig)$. Clearly, from Eq.~\eqref{eq:gc} no statement can
be inferred about $\hs$, which is a non-universal function containing the
details of the dynamics and the driving. On the other hand, the anti-symmetric
part is universal and originates from breaking time-reversal symmetry. As an
alternative explanation, Budini~\cite{budi11} has attributed the kink to
``intermittency'', i.e., the random switching between different dynamic
regimes. Here we show explicitly that the optimal trajectory minimizing the
stochastic action indeed corresponds to two different regimes depending on the
sign of $\sig$: forward ($\sig>0$) or backward ($\sig<0$). Considering only
the optimal trajectories, therefore, implies a non-differentiability in the
large deviation function at $\sig=0$, which, however, is smeared out by
fluctuations.


\section{Asymmetric random walk}

We first consider a particle on an infinite one-dimensional lattice. The
particle can hop to the left with rate $k^-$, or hop to the right with rate
$k^+$. A single trajectory of length $t$ is characterized by the number of
hops to the left, $n^-$, and to the right, $n^+$. Travelling the distance
$n\equiv n^+-n^-$, the entropy produced in the surrounding medium is
\begin{equation}
  \label{eq:arw:sm}
  \sm(n^+,n^-) = n\ln\frac{k^+}{k^-} \equiv nf,
\end{equation}
where
\begin{equation}
  \label{eq:arw:smm}
  \smm = f(k^+-k^-) = fk^+(1-e^{-f})
\end{equation}
is the mean entropy production rate. For $k^+=k^-$ the rates obey detailed
balance and clearly no entropy is produced on average.

\subsection{Donsker-Varadhan theory}

In a seminal paper on the fluctuation theorem for stochastic
dynamics~\cite{lebo99}, Lebowitz and Spohn have obtained the LDF for the
asymmetric random walk (ARW) from the cumulant generating function. The
normalized function $h(\sig)$ can be cast in the form
\begin{equation}
  \label{eq:arw:h}
  h(\sig) = (fz)^{-1}
  \left\{\cosh(f/2) +
    z\sig[\arsinh(z\sig)-f/2] - \sqrt{(z\sig)^2+1} \right\}
\end{equation}
with $z\equiv\sinh(f/2)$. It is straightforward to check that the function
$h(\sig)$ has the following properties: (i)~it is convex and non-negative,
$h(\sig)\geqslant0$, with minimum $h(1)=0$, (ii)~it obeys the Gallavotti-Cohen
symmetry Eq.~\eqref{eq:gc}, and (iii), following from this symmetry, $h(-1)=1$
holds.

\begin{figure}[t]
  \centering
  \includegraphics{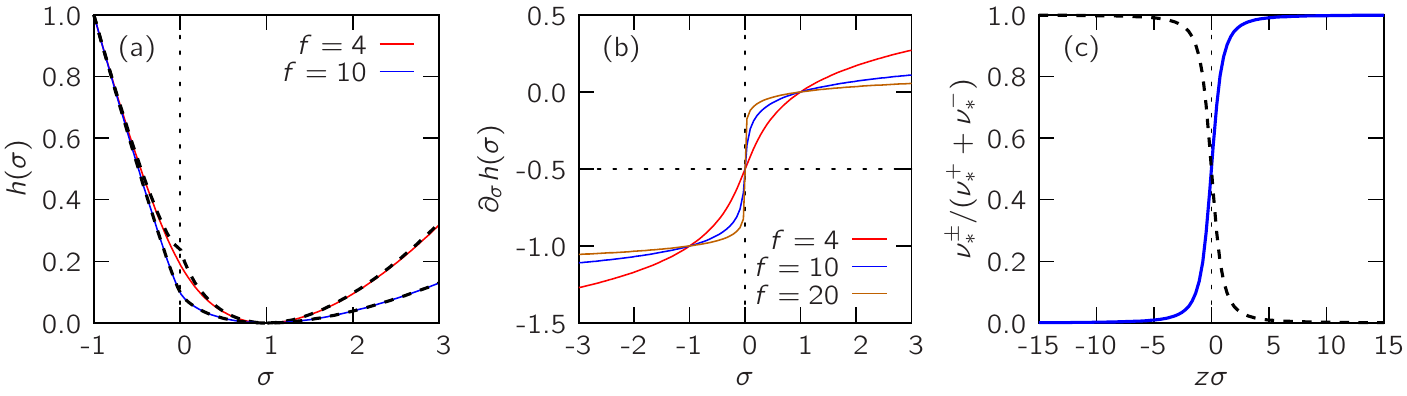}
  \caption{Asymmetric random walk: (a) Large deviation function $h(\sig)$ for
    two forces $f=4$ and $f=10$ (solid lines). Also shown is the limiting form
    Eq.~\eqref{eq:arw:hasym} (dashed lines). (b) The derivative becomes
    steeper at $\sig=0$ for higher driving force $f$. (c) The forward (solid
    line) and backward (dashed line) fluxes [Eq.~\eqref{eq:arw:flx}]
    \emph{vs.} the scaled rate $z\sig$, where $z=\sinh(f/2)$.}
  \label{fig:arw}
\end{figure}

To make our introductory comments concrete about a kink in the large deviation
function, for $|z\sig|\gg1$ we expand the inverse hyperbolic sine leading to
\begin{equation}
  h(\sig) \approx (fz)^{-1}\left\{\cosh(f/2)+z\sig[\pm\ln
    z|\sig|-f/2] - z|\sig|\right\}.
\end{equation}
The plus sign holds for $\sig>0$ while the minus sign holds for $\sig<0$. Here
and in the following we assume $f>0$. Considering the case $f\gg1$ of large
asymmetry we can further simplify
\begin{equation}
  \frac{\cosh(f/2)}{z} \approx 1,
  \qquad \ln z \approx -\frac{f}{2}
\end{equation}
yielding the limiting form
\begin{equation}
  \label{eq:arw:hasym}
  h(\sig) \approx \frac{1}{f}
  \left\{
    \begin{array}{ll}
      1+\sig(\ln\sig-1) & (\sig>0) \\
      1+|\sig|(\ln|\sig|-1+f) & (\sig<0)
    \end{array} \right.
\end{equation}
of the large deviation function in agreement with Ref.~\cite{doro11}. In
particular, the first derivative
\begin{equation}
  h'(\sig) \approx \frac{1}{f}
  \left\{
    \begin{array}{ll}
      \ln\sig & (\sig>0) \\
      -\ln|\sig| -f & (\sig<0)
    \end{array} \right.
\end{equation}
jumps at $\sig=0$. However, this jump is only apparent since for the expansion
leading to the limiting expression Eq.~\eqref{eq:arw:hasym} we used
$|\sig|\gg1/z$, i.e., we exclude a region of size $1/z$ around $\sig=0$. With
increasing $f$ this region becomes smaller but it does not vanish. Hence, for
any finite $f$ the function $h(\sig)$ is differentiable. In
Fig.~\ref{fig:arw}, the large deviation function and its first derivative are
plotted for different forces $f$.

\subsection{Alternative derivation}
\label{sec:arw:alt}

We now give an alternative derivation of Eq.~\eqref{eq:arw:h} in the spirit of
Freidlin-Wentzell. The path weight of a single trajectory is
\begin{equation}
  \label{eq:arw:P}
  P(n^+,n^-;t) = {m\choose n^+} \frac{t^m}{m!} (k^+)^{n^+}(k^-)^{n^-}
  e^{-(k^++k^-)t}
\end{equation}
with $m\equiv n^++n^-$ total jumps. This weight fulfills the normalization
condition
\begin{equation}
  \sum_{m=0}^\infty\sum_{n^+=0}^m P(n^+,n^-;t) = 1.
\end{equation}
Introducing dimensionless fluxes $\nu^\pm\equiv n^\pm/(k^+t)$ and expanding
the factorials using Stirling's approximation we obtain the rate function
\begin{eqnarray}
  \nonumber
  J(\nu^+,\nu^-) &\equiv& -\frac{1}{t}\ln P \\ &=&
  \label{eq:arw:ht}
  k^+[1 + e^{-f} - (\nu^+ + \nu^-) + f\nu^- +
  \nu^+\ln\nu^+ + \nu^-\ln\nu^-],
\end{eqnarray}
which does not depend on time explicitly. For long times $t$ the path weight
Eq.~\eqref{eq:arw:P} is sharply peaked around the minimum of $J$. We will call
trajectories that contribute to this minimum ``optimal'' trajectories. Note
that the small parameter $\eps$ in Eq.~\eqref{eq:fw} is now given by
$\eps=1/t$, i.e., here the time limits the fluctuations around the minimum.

For fixed $\sig$, forward and backward fluxes are not independent but related
through
\begin{equation}
  \label{eq:nu}
  \nu^+ = \nu^- + (1-e^{-f})\sig.
\end{equation}
From $\partial J/\partial\nu^-=0$ we then find that the minimum of
$J$ is reached for fluxes
\begin{equation}
  \label{eq:arw:flx}
  \nu_\ast^\pm = e^{-f/2}\left[\sqrt{(z\sig)^2+1} \pm z\sig \right], \qquad
  z\equiv\sinh(f/2).
\end{equation}
Using that
\begin{equation}
  \ln\nu_\ast^\pm = -\frac{f}{2}\pm\arsinh(z\sig)
\end{equation}
it is straightforward to show that $J(\nu_\ast^+,\nu_\ast^-)/\smm=h(\sig)$
yields the result Eq.~\eqref{eq:arw:h} obtained from the Donsker-Varadhan
theory.

The advantage of the alternative derivation is that it becomes transparent
which trajectories contribute to $h(\sig)$ for a given normalized entropy
production rate $\sig$. We now use this insight to discuss the physical origin
of the apparent kink. What does the optimal trajectory to reach a specified
entropy production, i.e., the trajectory that minimizes the rate function $J$,
look like? As becomes clear from Eq.~\eqref{eq:arw:flx}, for a given positive
$\sig\gg1/z$ the optimal trajectory consists almost entirely of forward
jumps. Conversely, for negative $\sig$ the optimal trajectory consists almost
entirely of backward jumps even though these trajectories are extremely
unlikely. This might seem surprising but any forward jump has to be
compensated by an unlikely backward jump. Setting $n^-=0$ for $\sig>0$ and
$n^+=0$ for $\sig<0$ leads to the two branches of the limiting form given in
Eq.~\eqref{eq:arw:hasym}. The reason why nevertheless $h(\sig)$ is
differentiable is that for $|\sig|\lesssim 1/z$ many trajectories contribute
to the large deviation function. In particular, for $\sig=0$ the forward and
backward fluxes become equal, $\nu^+=\nu^-$. All trajectories with the same
number of forward and backward jumps contribute since they have the same
weight. Their number is given by the binomial coefficient in
Eq.~\eqref{eq:arw:P}. In contrast, there is only one optimal combination
$(n^+,n^-)$ in the wings of the distribution.


\section{Driven Brownian motion}

In this section we consider a single colloidal particle moving in a closed
ring geometry with circumference $L$ and external periodic potential
$V(x+L)=V(x)$. The overdamped dynamics are described by the Langevin equation
\begin{equation}
  \label{eq:lang}
  \dot x = \mu_0[-V'(x)+f] + \eta,
\end{equation}
where the prime denotes the derivative with respect to $x$. The particle is
driven out of equilibrium through the force $f$ resulting in a non-vanishing
current through the ring. The noise $\eta$ describes the effective interactions
between the solvated particle and the solvent at temperature $T$ with
correlations
\begin{equation}
  \label{eq:corr}
  \mean{\eta(\tau)\eta(\tau')} = 2D_0\delta(\tau-\tau').
\end{equation}
Through $D_0=\mu_0 T$ we assume that the solvent remains at equilibrium even
though the particle is driven. To simplify the notation we will use
dimensionless quantities
\begin{equation}
  x \mapsto Lx, \quad V(x) \mapsto TV(x), \quad f \mapsto \frac{T}{L}f,
  \quad \tau \mapsto \frac{L^2}{D_0}\tau
\end{equation}
throughout the remainder of this paper. Moreover, these units make transparent
the connection with the ARW. In particular, the number of periods traversed is
$n$ and the dimensionless force $f$ corresponds to the force defined in
Eq.~\eqref{eq:arw:sm}. We will also need the notion of a critical force $\fc$
such that for $f>\fc$ the deterministic force $-V'(x)+f>0$ is positive for all
$x$ and, therefore, in the absence of noise running solutions exist.

The rate of entropy production is ${\dot s}_\mathrm{m}=f\dot x$ with
$\sm=f\Delta x$ [cf. Eq.~(\ref{eq:arw:sm}) for the ARW], where $\Delta x$ is
the distance the particle has traveled during the time $t$. The mean velocity
of the particle can be calculated from Stratonovich's formula~\cite{risken}
\begin{equation}
  \vel = \kap(1-e^{-f})
\end{equation}
with
\begin{eqnarray}
  \kap^{-1} \equiv
  \IInt{x}{0}{1}\IInt{z}{0}{1} \exp\left\{V(x)-V(x-z)-fz\right\}.
\end{eqnarray}
The mean velocity (and also the mean entropy production) of ARW and ring
potential agree if we set $k^+=\kap$, cf. Eq.~\eqref{eq:arw:smm}. However, in
general fluctuations will be different, leading to different shapes for the
LDFs. For large forces $f$ the influence of the potential on the particle
motion becomes negligible and $\vel\approx f$.

\subsection{Donsker-Varadhan theory}

The time evolution of the probability distribution of $x$ is governed through
the Fokker-Planck operator
\begin{equation}
  \label{eq:fp}
  \hat L_0 \equiv \pd{}{x}\left[(V'-f) + \pd{}{x}\right].
\end{equation}
In particular, the steady state distribution obeys $\hat
L_0p_\mathrm{ss}(x)=0$.  To calculate $\al(\lam)$ [Eq.~\eqref{eq:al}] consider
the two-sided Laplace transform
\begin{equation}
  \label{eq:ld:joint}
  \tilde p(x,\lam,t) = \IInt{\sm}{-\infty}{+\infty} e^{\lam\sm} p(x,\sm,t)
\end{equation}
of the joint probability $p(x,\sm,t)$ to find the particle at position $x$
with an amount of entropy $\sm$ produced up to time $t$. The time evolution of
the transformed joint probability obeys $\partial_t\tilde p=\hat L_\lam\tilde
p$, where the operator
\begin{equation}
  \label{eq:OL}
  \hat L_\lam = \hat L_0 + \left[2\pd{}{x}+V'-f\right](f\lam) + (f\lam)^2
\end{equation}
has been determined in Ref.~\cite{mehl08}. We expand the joint probability
in eigenfunctions of the time evolution operator
\begin{equation}
  \tilde p(x,\lam,t) = \sum_{i=0}^\infty c_i e^{\al_i(\lam)t}\psi_i(x,\lam),
\end{equation}
where the coefficients $c_i$ quantify the overlap with the initial
distribution $\tilde p(x,\lam,0)=p_\mathrm{ss}(x)$. The eigenvalue equations
read $\hat L_\lam\psi_i(x,\lam)=\al_i(\lam)\psi_i(x,\lam)$. In the limit of
long times, the joint probability is dominated by the largest eigenvalue
$\al_0$. Integrating Eq.~\eqref{eq:ld:joint} over the position $x$ thus leads
to
\begin{equation}
  \mean{e^{\lam\sm}} \sim e^{\al_0(\lam)t},
\end{equation}
and we conclude that the scaled cumulant generating function
$\al(\lam)=\al_0(\lam)$ is indeed given by the largest eigenvalue of the time
evolution operator~(\ref{eq:OL}). It can be calculated solving either the
eigenvalue problem directly~\cite{mehl08} or through the Ritz variational
method~\cite{laco09}. The function $h(\sig)$ then follows
from~\eqref{eq:legendre}. Here we use the first method and refer to
Ref.~\cite{mehl08} for further details.

\subsection{Stochastic action and effective potential}

To continue we employ the well-established machinery of path integrals for
stochastic processes~\cite{kleinert} that has been developed to deal, e.g.,
with diffusion in bistable potentials~\cite{caro81} and escape
problems~\cite{lehm00}. Imagine that the particle starts at position $x_0$ and
travels a distance $\Delta x$ during the time $t$. The probability for such a
transition
\begin{equation}
  \label{eq:P}
  P(x_0+\Delta x|x_0;t) = \int_{x_0}^{x_0+\Delta x}[\dd x(\tau)]
  e^{-\ac[x(\tau)]}
\end{equation}
is obtained through summing over all possible paths with stochastic action
\begin{equation}
  \label{eq:S}
  \ac[x(\tau)] = \IInt{\tau}{0}{t} \left\{\frac{1}{4}(\dot x+V'-f)^2 -
    \frac{1}{2}V'' \right\}
\end{equation}
calculated along a single path $x(\tau)$. The last term $V''/2$ is the
Jacobian arising from the change of variables $\eta\ra x$. 

The stochastic action Eq.~\eqref{eq:S} can be rearranged to give
\begin{equation}
  \label{eq:S:L}
  \ac[x(\tau)] = \frac{1}{2}\Delta V - \frac{1}{2}f\Delta x +
  \frac{1}{4} f^2t + \IInt{\tau}{0}{t} \la(\tau).
\end{equation}
The last term 
\begin{equation}
  \label{eq:L}
  \la(x,\dot x) = \frac{1}{4}\dot x^2 - U(x)
\end{equation}
takes on the form of a Lagrangian for a classical particle with mass $m=1/2$
moving in the effective potential
\begin{equation}
  \label{eq:U}
  U(x) \equiv \frac{1}{2}V''(x) - \frac{1}{4}[V'(x)]^2 + \frac{1}{2}fV'(x).
\end{equation}
It is straightforward to check that for a periodic (and twice differentiable)
potential $V(x)$ the effective potential $U(x)$ is periodic, too. For
completeness, we note that the same effective potential arises in the
symmetrization of the Fokker-Planck operator [Eq.~\eqref{eq:fp}],
\begin{equation}
  e^{V(x)/2}\hat L_0e^{-V(x)/2} = \pd{^2}{x^2} + U(x),
\end{equation}
see, e.g., the textbook treatment in Ref.~\cite{risken}.

\begin{figure}[b!]
  \centering
  \includegraphics{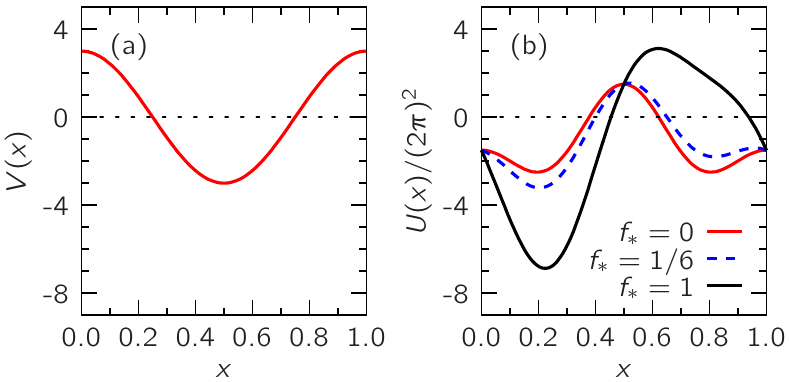}
  \caption{(a)~Original potential $V(x)=v_0\cos(2\pi x)$ and (b)~the mapped
    effective potential $U(x)$ from Eq.~\eqref{eq:U} plotted for $v_0=3$ and
    forces $f=0$, $f<\fc$, and at the critical force $f=\fc=6\pi$.}
  \label{fig:sketch}
\end{figure}

In Fig.~\ref{fig:sketch} the effective potential $U(x)$ is plotted for the
cosine potential $V(x)=v_0\cos(2\pi x)$ with critical force $\fc=2\pi v_0$. It
will turn out useful to define the reduced force $f_\ast\equiv f/\fc$. In the
following we will use this specific potential for all illustrations. In
equilibrium, the principal feature of the effective potential is that the
minimum of $V(x)$ becomes the global maximum of $U(x)$. For $v_0>1$, the
minimum of $U(x)$ splits into two minima, which are degenerate at equilibrium
($f=0$). Turning on the driving force ($f>0$) the global maximum of $U(x)$ is
shifted, the depth of the left minimum increases, and the depth of the right
minimum decreases until for forces $f_\ast\geqslant(1-v_0^{-2/3})^{3/2}$ only
one global minimum remains. For $f_\ast\geqslant1$ the upper arc becomes
concave.

\subsection{Optimal trajectory}

We now determine the trajectory $x_\ast(\tau)$ that minimizes the action
[Eq.~\eqref{eq:S:L}] with $\ac_\ast\equiv\ac[x_\ast(\tau)]$. Clearly, this
trajectory is the solution of the Euler-Lagrange equation
\begin{equation}
  \label{eq:EL}
  -\fd{\ac}{x(\tau)} = \td{}{\tau}\pd{\la}{\dot x} - \pd{\la}{x}
  = \frac{1}{2}\ddot x + U'(x) = 0
\end{equation}
subject to the boundary conditions $x_\ast(0)=x_0$ and $x_\ast(t)=x_0+\Delta
x$. Since the effective potential is periodic, the solution
$x_\ast(\tau+\tp)=x_\ast(\tau)$ of the Euler-Lagrange Eq.~\eqref{eq:EL} will
also be periodic with period
\begin{equation}
  \tp = \frac{1}{|\sig|\mean{\dot x}}.
\end{equation}
The sign of $\sig$ determines the sign of the initial velocity of the
particle. Note that the solution of Eq.~\eqref{eq:EL} is time-reversal
symmetric.

Before we proceed we shift the effective potential
\begin{equation}
  U(x) = U_\ast + W(x), \qquad U_\ast = \max_x U(x)
\end{equation}
such that $W(x)\leqslant0$ is non-positive. Although no small parameter
appears in the stochastic action, we nevertheless follow the Freidlin-Wentzell
route and assume that the transition probability is dominated by the optimal
trajectory. We thus obtain from Eq.~\eqref{eq:S:L} the unnormalized, naive
approximation
\begin{equation}
  \label{eq:h:0}
  \tilde h_\ast(\sig) \equiv \lim_{t\ra\infty}\frac{\ac_\ast}{t}
  = -\frac{1}{2}\sig\smm + \frac{1}{4}f^2 + \frac{A}{\tp} - U_\ast
\end{equation}
with optimal stochastic action for a single period
\begin{equation}
  \label{eq:A}
  A \equiv \IInt{\tau}{0}{\tp} \left\{ \frac{1}{4}[\dot x_\ast(\tau)]^2 -
    W(x_\ast(\tau)) \right\}.
\end{equation}
Since the potential $V(x)$ is bounded, the term $\Delta V/t$ drops out in the
long time limit. We stress that, since no small parameter appears in the
stochastic action, it cannot be expected in general that $h_\ast(\sig)$
corresponds to $h(\sig)$ as obtained by the operator method.

\begin{figure}[t]
  \centering
  \includegraphics{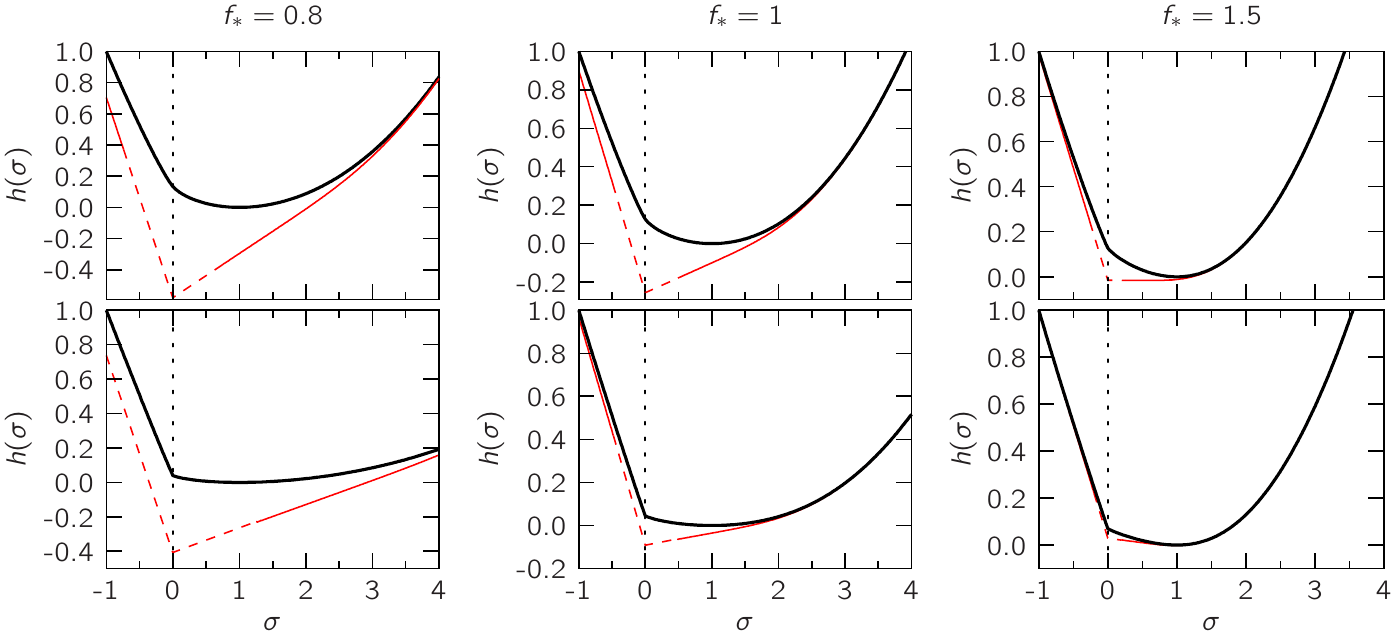}
  \caption{Large deviation function for the periodic potential
    $V(x)=v_0\cos(2\pi x)$ with $v_0=2$ (top row) and $v_0=6$ (bottom row) for
    three reduced driving forces $f_\ast$. Shown are the exact solutions
    $h(\sig)$ (thick lines) obtained from the operator method together with
    the naive approximations $h_\ast(\sig)$ [Eq.~\eqref{eq:h:norm}, thin
    lines]. For small $\sig$ the numerical calculation breaks down and the
    dashed lines show the linear extrapolation towards the known value
    $h_\ast(0)$ [see Eq.~\eqref{eq:h0}].}
  \label{fig:ring}
\end{figure}

The time integral in Eq.~\eqref{eq:A} is evaluated most conveniently by
changing the integration variable from $\tau$ to $x$. To this end we note that
\begin{equation}
  \label{eq:E}
  E = \frac{1}{4}[\dot x_\ast(\tau)]^2 + W(x_\ast(\tau))
\end{equation}
is a constant of motion. The time to traverse a single period is
\begin{equation}
  \label{eq:tau0}
  \tp(E) = \frac{1}{2} \IInt{x}{0}{1} [E-W(x)]^{-1/2}.
\end{equation}
Rearranging Eq.~\eqref{eq:A} using Eq.~\eqref{eq:E}, the action for a single
period reads
\begin{equation}
  \label{eq:A:E}
  A(E) = \IInt{x}{0}{1} \frac{E-2W(x)}{2\sqrt{E-W(x)}} 
  = \IInt{x}{0}{1} \sqrt{E-W(x)} - E\tp(E).
\end{equation}
We thus solve Eq.~\eqref{eq:tau0} for a given time $\tp$ in order to obtain
$E$, from which we determine the optimal action $A$. The resulting normalized
function
\begin{equation}
  \label{eq:h:norm}
  h_\ast(\sig)=\tilde h_\ast(\sig)/\smm
\end{equation}
is plotted in Fig.~\ref{fig:ring} for different driving forces $f$ below and
above the critical force $\fc$. The calculation breaks down for small $\sig$
as the initial velocity, and therefore $E$, become of the order of the machine
precision. We find that for large $\sig$ the optimal path solution agrees with
the full solution $h(\sig)$ but deviates for small $\sig$, where the
disagreement becomes more pronounced for small driving forces.

We now consider two special cases. For large $E$ the influence of the
potential is negligible and the velocity $\dot x_\ast\approx\pm2\sqrt{E}$ is
approximately constant, with $\dot x_\ast=\pm1/\tp=\sig\vel$. Together with
$A=E\tp$ we obtain from Eq.~\eqref{eq:h:0}
\begin{equation}
  \tilde h_\ast(\sig) = \frac{1}{4}(\sig\vel-f)^2 - U_\ast,
\end{equation}
i.e., the asymptotic form of the large deviation function for large rates
$|\sig|$ is a parabola. In particular, for a flat potential $V(x)=0$ with
$\vel=f$ we recover the exact result $h(\sig)=(1/4)(\sig-1)^2$.

\begin{figure}[t]
  \centering
  \includegraphics{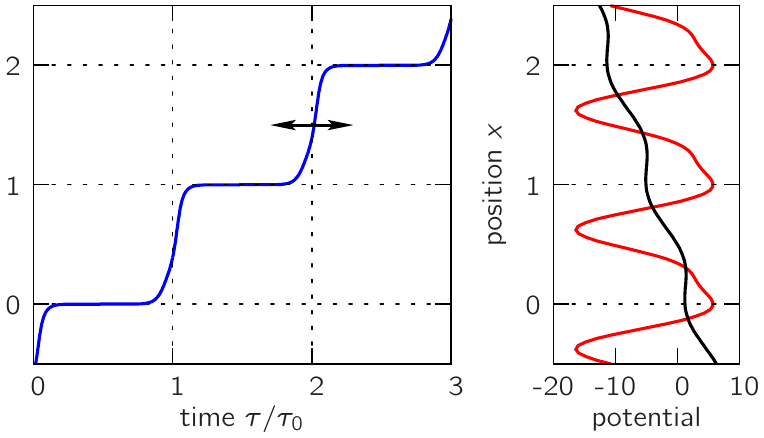}
  \caption{Optimal trajectory $x_\ast(\tau)$ plotted over three periods
    (parameters: $v_0=5$, $f=8\pi$, $\sig=1$). The potentials $U(x)$
    (red/bright) and $V(x)-fx$ (black) are shown in the right panel.}
  \label{fig:traj}
\end{figure}

For $E\ra0$ the period $\tp$ increases dramatically and the particle spends a
long time at the barriers of the effective potential [corresponding to the
valleys of the original potential $V(x)$], see Fig.~\ref{fig:traj}. Up to
linear order in $E$ the optimal action Eq.~\eqref{eq:A:E} becomes
\begin{equation}
  A \simeq A_0 \equiv A(0) = \IInt{x}{0}{1} \sqrt{-W(x)}.
\end{equation}
This is of course the action of an \emph{instanton}, the classical solution
for $E=0$ that connects two adjacent barriers in infinite time. To
approximately calculate the period $\tp$ we expand $W(x)$ to second order,
\begin{equation}
  \frac{\tp}{2} \approx \frac{1}{2}\IInt{x}{0}{\frac{1}{2}}
  [E-U_\mathrm{b}''x^2/2]^{-1/2},
\end{equation}
where, without loss of generality, we have shifted the potential along the
$x$-axis such that $W(0)=0$, and $U_\mathrm{b}''<0$ is the curvature of the
effective potential at the barrier. To leading order we find the well-known
result~\cite{caro81}
\begin{equation}
  \label{eq:E:small}
  E \simeq \frac{|U_\mathrm{b}''|}{2}
  \exp\left\{-\sqrt{2|U_\mathrm{b}''|}\tp\right\}.
\end{equation}
While the time $\tp$ to traverse a single period diverges for $\sig\ra0$, the
action of an instanton remains finite and $A_0/\tp\ra0$. The approximated
large deviation function at vanishing entropy production $\sig=0$ thus reads
\begin{equation}
  \label{eq:h0}
  \tilde h_\ast(0) = \frac{1}{4}f^2 - U_\ast.
\end{equation}
The linear extrapolation of the numerically determined $h_\ast(\sig)$ towards
this value is shown in Fig.~\ref{fig:ring}.

Summarizing, the results of naively applying the Freidlin-Wentzell approach to
the calculation of the large deviation function are: (i)~The optimal path is
not sufficient to reproduce the large deviation function over the full range
of rates $\sig$. (ii)~Only accounting for the contribution from the optimal
path results in a function $h_\ast(\sig)$ that is non-differentiable at
$\sig=0$, i.e., the approximated large deviation function exhibits a genuine
kink. (iii)~The Gallavotti-Cohen symmetry~\eqref{eq:gc} is fulfilled by both
$h_\ast(\sig)$ and $h(\sig)$. The rounding of the kink in the full solution
$h(\sig)$ points to the importance of fluctuations around the optimal path
even in the limit $t\ra\infty$.

\subsection{Small-noise limit}

As mentioned in the introduction, the Freidlin-Wentzell method is strictly
valid only in the limit $\eps\ra0$ of vanishing noise. In our dimensionless
units, the small parameter $\eps=1/v_0$ is the ratio of thermal energy to the
potential depth. Fixing $f_\ast$ with $\fc\propto v_0$ we see that the first
term in Eq.~\eqref{eq:U} is of order $v_0$ and the other two terms are of
order $v_0^2$. We thus drop the first term (the Jacobian). Rearranging terms
we obtain for the effective potential the simplified expression
\begin{equation}
  U(x) = \frac{1}{4}f^2 - \frac{1}{4}[-V'(x)+f]^2.
\end{equation}
In the following we focus on the case $f<\fc$. The maximum of the potential is
then reached at points $x$ where the deterministic force $-V'(x)+f$ is zero
with
\begin{equation}
  U_\ast = \frac{1}{4}f^2 \qquad (f<\fc).
\end{equation}
From Eq.~\eqref{eq:h0} we then know that $h_\ast(0)=0$. Since $h_\ast(\sig)$
is a convex function with $h_\ast(1)=0$ we can infer that the shape in the
range $-1\leqslant\sig\leqslant1$ is a ``wedge''. The approximated LDF is
universal in this range and does not depend on $f$ (as long as $f<\fc$) nor
$v_0$.

\begin{figure}[t]
  \centering
  \includegraphics{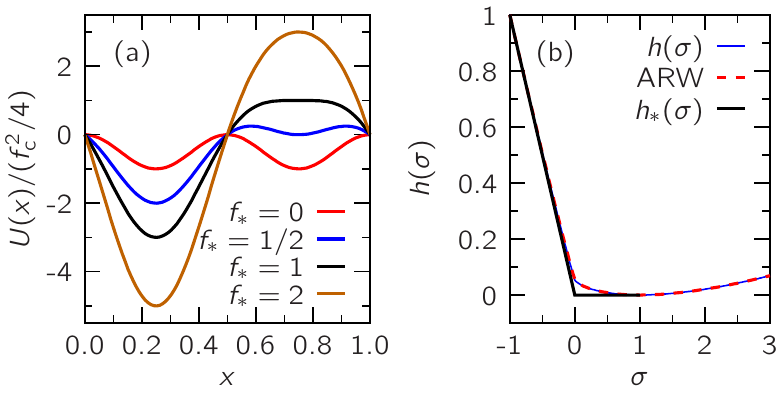}
  \caption{Small-noise limit: (a)~Simplified effective potential $U(x)$
    dropping the Jacobian in Eq.~\eqref{eq:U} [cf.
    Fig.~\ref{fig:sketch}(b)]. (b)~Approximated LDF $h_\ast(\sig)$ for
    $f_\ast<1$ (thick line). Also shown is the LDF $h(\sig)$ for $v_0=7$ and
    $f=6\pi$ ($f_\ast=3/7$) together with the ARW solution
    Eq.~\eqref{eq:arw:h} for the same force.}
  \label{fig:smnoise}
\end{figure}

Specifically, for the cosine potential $V(x)=v_0\cos(2\pi x)$ the effective
potential
\begin{equation}
  U(x) = \frac{\fc^2}{4}\left\{ f_\ast^2 - [\sin(2\pi x)+f_\ast]^2 \right\}
\end{equation}
is shown in Fig.~\ref{fig:smnoise}(a), whereas the ``wedge''-shaped
$h_\ast(\sig)$ is depicted in Fig.~\ref{fig:smnoise}(b). The action of an
instanton is easily evaluated to give
\begin{equation}
  \label{eq:sm:A}
  A_0 = \IInt{x}{0}{1} \frac{\fc}{2}\left|\sin(2\pi x)+f_\ast\right|
  = \frac{\fc}{\pi}\left[\sqrt{1-f_\ast^2}+f_\ast\arcsin f_\ast\right].
\end{equation}
Also shown in Fig.~\ref{fig:smnoise}(b) is that the actual LDF $h(\sig)$
agrees with the ARW solution for large $v_0$. Since $f_\ast$ is fixed the
force $f$ increases as we increase the potential depth $v_0$. Both the ARW and
the actual solution $h(\sig)$ then approach the limiting ``wedge'' shape as
$v_0\ra\infty$.


\subsection{Role of fluctuations}

At this point it seems natural to inquire what the nature is of the
fluctuations contributing to the large deviation function. Specifically, we
consider two types of fluctuations: small Gaussian perturbations around the
optimal path, and fluctuations of the ``jumps'' times $\tau_l$. In the
periodic optimal trajectory $x_\ast(\tau)$ these jumps occur regularly with
period $\tp$, see Fig.~\ref{fig:traj}.

\subsubsection{Gaussian fluctuations.}

For finite $t$, we expand the action Eq.~\eqref{eq:S:L}
\begin{equation}
  \label{eq:S:Q}
  \ac[x_\ast(\tau)+\xi(\tau)] \approx \ac_\ast 
  + \frac{1}{2} \IInt{\tau}{0}{t} \xi(\tau)\hat D(\tau)\xi(\tau)
\end{equation}
to second order with small perturbations $\xi(\tau)$ of the optimal path,
which obey the boundary conditions $\xi(0)=\xi(t)=0$. Here we have introduced
the symmetric second variation operator
\begin{equation}
  \label{eq:SL}
  \hat D(\tau) \equiv -\frac{1}{2}\td{^2}{\tau^2} - U''(x_\ast(\tau)),
\end{equation}
which is formally equivalent to a Schr\"odinger equation with potential energy
$-U''(x_\ast(\tau))$ (positive curvatures of the effective potential
correspond to low ``energies''). We, therefore, can immediately state that the
eigenvalues $\lam_i$ of $\hat D$ are real and can be ordered,
$\lam_0<\lam_1<\cdots$. Moreover, eigenfunctions $\xi_i(\tau)$ corresponding
to different eigenvalues $\lam_i$ are orthogonal,
\begin{equation}
  \label{eq:xi}
  \IInt{\tau}{0}{t} \xi_i(\tau)\xi_j(\tau) = \delta_{ij}, \qquad
  \xi_i(0) = \xi_i(t) = 0.
\end{equation}
We expand the generic perturbation $\xi(\tau)$ in the basis spanned by the
normalized eigenfunctions $\{\xi_i\}$,
\begin{equation}
  \label{eq:pert}
  \xi(\tau) = \sum_{i=1}^\infty c_i\xi_i(\tau),
\end{equation}
with coefficients $c_i$. The path measure is
\begin{equation}
  [\dd x(\tau)] \propto \prod_{i=1}^\infty \dd c_i
\end{equation}
and we perform the Gaussian integration of Eq.~\eqref{eq:P} with result
\begin{equation}
  \label{eq:P:gauss}
  P(x_0+\Delta x|x_0;t) \simeq \frac{e^{-\ac_\ast}}{\sqrt{\det\hat D}},
  \qquad
  \det\hat D=\prod_{i=1}^\infty\lam_i.
\end{equation}
Of course, this result only holds if all eigenvalues are positive. We
numerically calculate the determinant using Floquet theory as outlined in the
appendix. We find that small Gaussian fluctuation around the optimal path
$x_\ast(\tau)$ do not contribute to the large deviation function.

\subsubsection{Dilute instanton gas.}

For $f\ll\fc$ the particle in the periodic continuous potential $V(x)$
effectively undergoes a hopping motion resembling the discrete random
walk. For such thermally activated jumps, the corresponding rate is
\begin{equation}
  k^+ \approx \exp\left\{-\Delta V + f(x^+-x^-) \right\},
\end{equation}
where $x^-$ is the position of the minimum, $x^+$ the position of the barrier,
and $\Delta V\equiv V(x^+)-V(x^-)$ is the barrier height. For the cosine
potential we find
\begin{equation}
  2\pi x^- = \pi + \arcsin f_\ast, \quad
  2\pi x^+ = 2\pi - \arcsin f_\ast.
\end{equation}
The rate thus reads
\begin{equation}
  \label{eq:sm:k}
  k^+ \approx e^{-A_0+f/2},
\end{equation}
where $A_0$ is the action [Eq.~\eqref{eq:sm:A}] of an instanton in the small
noise limit and we have ignored a sub-exponential prefactor.

The optimal trajectory $x_\ast(\tau)$ is strictly periodic. However,
Fig.~\ref{fig:traj} suggests that shifting the positions of the ``jumps''
$\tau_l=l\tp$ does only minimally change the action as long as the particle
spends a quasi-infinite time on top of the barrier (of the effective potential
$U$). In this case some eigenvalues $\lam_i$ become exponentially small (and
therefore contribute to the large deviation function) or even vanish, in which
case the calculation of the previous subsection breaks down. Such zero-modes
need to be treated in a different way known as the method of collective
coordinates. In effect, integrations over zero-modes are replaced by
integrations over the corresponding jump time.

Suppose a trajectory is composed of $n^+$ instantons and $n^-$ anti-instantons
jumping against the driving force. We use again $n=n^+-n^-$ and
$m=n^++n^-$. If we neglect all ``interactions'' between instantons we can
approximate the action as
\begin{equation}
  \ac \approx -\frac{1}{2}fn + mA_0
\end{equation}
independent of the actual jump times. Integration over the jump times then
leads to a path weight
\begin{equation}
  P(n^+,n^-;t) = \frac{1}{Z}{m\choose n^+} \frac{t^m}{m!} e^{-\ac}
\end{equation}
with normalization factor
\begin{equation}
  Z = \exp\left\{2t\cosh(f/2)e^{-A_0}\right\}.
\end{equation}
We now follow the alternative derivation for the asymmetric random walk
presented in Sec.~\ref{sec:arw:alt}. Introducing dimensionless fluxes
$\nu^\pm\equiv n^\pm/(\kap t)$ [related by Eq.~\eqref{eq:nu}], we obtain the
rate function
\begin{eqnarray}
  \nonumber
  J(\nu^+,\nu^-) &=& 2\cosh(f/2)e^{-A_0} + \kap\left\{\nu^+\ln\nu^+ +
    \nu^-\ln\nu^- \right. \\ && \left.
    - (f/2)(\nu^+-\nu^-) - (\nu^++\nu^-)(1-A_0-\ln\kap) \right\}.
\end{eqnarray}
The minimum is attained at fluxes
\begin{equation}
  \nu^\pm_\ast = \frac{e^{-A_0}}{\kap}
  \left[\sqrt{(\zeta\sig)^2+1} \pm \zeta\sig\right]
\end{equation}
with
\begin{equation}
  \zeta \equiv \frac{1}{2}\kap e^{A_0}(1-e^{-f}).
\end{equation}
Plugging in the rate Eq.~\eqref{eq:sm:k} for $\kap=k^+$ we find $\zeta\approx
z$ and, moreover, we recover the form Eq.~\eqref{eq:arw:h} of the LDF for the
ARW.


\section{Conclusions}

We have studied the large deviation function for the entropy production in two
simple one-dimensional systems: the asymmetric random walk of a particle on a
discrete lattice and the continuous motion of a driven particle in an external
potential. For both systems we have calculated the large deviation function
using the path approach of Freidlin and Wentzell. However, while for the ARW
the solution thus obtained agrees with the Donsker-Varadhan theory, for the
continuous case the solution is only an approximation. In both approaches, the
wings of the large deviation function are well described by a single
trajectory composed of only forward ($\sig>0$) or only backward ($\sig<0$)
jumps. Changing the prescribed rate $\sig$, the transition between these two
regimes appears as a kink at $\sig=0$, which in the true large deviation
function is smeared out by fluctuations. For the ARW the reason is that there
are many combinations of forward and backward jumps (and therefore many
trajectories) around $\sig=0$ as described by the combinatorial factor in
Eq.~\eqref{eq:arw:P}. For the continuous system we have verified numerically
that small perturbations of the optimal path do not contribute to the large
deviation function. For large $v_0$ and small driving forces $f_\ast\ll1$
trajectories are dominated by discrete barrier crossings and can be decomposed
into instanton solutions. Summing these trajectories, we recover the large
deviation function of the ARW.


\ack

We thank Hugo Touchette for valuable discussions. TS gratefully acknowledges
financial support by the Alexander-von-Humboldt foundation during the initial
stage of this work.


\appendix

\section{Floquet theory}

We estimate the effect of small fluctuations through calculating the full
determinant $\det\hat D$ of the operator Eq.~\eqref{eq:SL}. To this end we
make use of the following relation
\begin{equation}
  \frac{\det\hat D}{\det\hat D_0} = \frac{\phi(t)}{\phi_0(t)}
\end{equation}
with $\hat D_0$ a suitable reference operator~\cite{kirs03}. In the simplest
case $\hat D_0=-\partial_\tau^2$ but the exact form of $\hat D_0$ does not
play a role. Here, $\phi(t)$ is the solution of $\hat D\phi=0$ whereas
$\phi_0(t)$ is the solution of $\hat D_0\phi_0=0$. Defining the vector $\vec
x\equiv(\phi,\dot\phi)$ we obtain a linear, periodic differential equation
\begin{equation}
  \label{eq:floq}
  \dot{\vec x} = \mat M(x_\ast(\tau))\vec x, \qquad
  \mat M(x) \equiv \left(
    \begin{array}{cc}
      0 & 1 \\ -2U''(x) & 0
    \end{array}\right).
\end{equation}
From Floquet theory it is well-known that the solution can be written
\begin{equation}
  \label{eq:floq:sol}
  \vec x(\tau) = c_+e^{+\mu\tau}\vec p(\tau) + c_-e^{-\mu\tau}\vec p(-\tau)
\end{equation}
with characteristic, or Floquet, exponent $\mu$. Hence, in the limit of long
times the determinant behaves as $\det\hat D\sim\phi(t)\sim|e^{\mu t}|$ to
leading order. We again change variables from $\tau$ to $x$ and solve the
ordinary differential equation
\begin{equation*}
  \pd{\mat X}{x} = \frac{1}{\dot x_\ast}\mat M(x)\mat X =
  \frac{1}{2}[E-U(x)]^{-1/2} \mat M(x)\mat X
\end{equation*}
for a matrix $\mat X$ with the matrix $\mat M(x)$ given in
Eq.~\eqref{eq:floq}. The initial condition is $\mat X(0)=\mathbf 1$, and we
integrate a single period. The resulting matrix $\mat X(1)$ has two
eigenvalues $\rho_\pm=e^{\pm\mu}$. Numerically we find $|\rho_\pm|\simeq1$ for
the values of $E$ that can be obtained numerically. Therefore, small Gaussian
fluctuations around the optimal path $x_\ast(\tau)$ do not contribute to the
large deviation function in the long time limit.


\end{document}